\documentclass[runningheads]{llncs}
\usepackage[T1]{fontenc}

\usepackage{svg}
\usepackage{amsmath}
\usepackage{tikz}
\usepackage{graphicx}
\usepackage{caption}

\usepackage{amsfonts}
\usepackage{mathtools}
\usepackage{framed}
\usepackage{microtype}
\usepackage{adjustbox}
\DeclareMathOperator*{\expectation}{\mathbb{E}}

\usepackage{subcaption}
\usepackage{epsfig}
\usepackage{url}

\usepackage{lipsum}

\usepackage{adjustbox}
\usepackage{floatrow}
\usepackage{graphics}

\usepackage[font=small,labelfont=bf]{caption}

\usepackage{graphicx}
\usepackage{xcolor}
\usepackage{array}
\usepackage{amsmath}
\usepackage{amsfonts}
\usepackage{soul}
\begin{document}
\title{LMT: Longitudinal Mixing Training, a framework to predict disease progression from a single image}
%

\titlerunning{LMT: Longitudinal Mixing Training to predict disease progression}

%
\author{Rachid Zeghlache\inst{1,2}
\and
Pierre-Henri Conze\inst{1,3} 
\and
Mostafa El Habib Daho \inst{1,2}
\and
Yihao Li \inst{1,2}
\and
Hugo Le boite \inst{5}
\and
Ramin Tadayoni\inst{5} 
\and
Pascal Massin\inst{5} 
\and
Béatrice Cochener\inst{1,2,4} 
\and
Ikram Brahim \inst{1,6} 
\and 
Gwenolé Quellec\inst{1} 
\and
Mathieu Lamard\inst{1,2} 
}

\institute{
LaTIM UMR 1101, Inserm, Brest, France \and
University of Western Brittany, Brest, France \and
IMT Atlantique, Brest, France
\and
Ophtalmology Department, CHRU Brest, Brest, France \and
Lariboisière Hospital, AP-HP, Paris, France\and
LBAI UMR 1227, Inserm, Brest, France
}

\authorrunning{R. Zeghlache et al.}

\maketitle              
\begin{abstract}

Longitudinal imaging is able to capture both static anatomical structures and dynamic changes in disease progression toward earlier and better patient-specific pathology management. However, conventional approaches rarely take advantage of longitudinal information for detection and prediction purposes, especially for Diabetic Retinopathy (DR). In the past years, Mix-up training and pretext tasks with longitudinal context have effectively enhanced DR classification results and captured disease progression. In the meantime, a novel type of neural network named Neural Ordinary Differential Equation (NODE) has been proposed for solving ordinary differential equations, with a neural network treated as a black box. By definition, NODE is well suited for solving time-related problems. In this paper, we propose to combine these three aspects to detect and predict DR progression. Our framework, Longitudinal Mixing Training (LMT), can be considered both as a regularizer and as a pretext task that encodes the disease progression in the latent space.  Additionally, we evaluate the trained model weights on a downstream task with a longitudinal context using standard and longitudinal pretext tasks. We introduce a new way to train time-aware models using $t_{mix}$, a weighted average time between two consecutive examinations. We compare our approach to standard mixing training on DR classification using OPHDIAT a longitudinal retinal Color Fundus Photographs (CFP) dataset. We were able to predict whether an eye would develop a severe DR in the following visit using a single image, with an AUC of 0.798 compared to baseline results of 0.641. Our results indicate that our longitudinal pretext task can learn the progression of DR disease and that introducing $t_{mix}$ augmentation is beneficial for time-aware models.

\end{abstract}

\section{Introduction}

According to the International Diabetes Federation, by 2045, diabetes will impact 700 million individuals globally, with over one-third suffering from Diabetic Retinopathy (DR) \cite{Saeedi2019}. DR is the leading cause of vision loss worldwide and is caused by high blood sugar damaging retinal vessels, leading to swelling and leakage \cite{ogurtsova2017idf}. Color fundus photographs (CFP) are used clinically to detect DR. Severity is graded in five classes (0, 1, 2, 3 and 4) using the International Clinical DR (ICDR) scale: 0 is no apparent DR, 1 is mild non-proliferative DR (NPDR), 2 is moderate NPDR, 3 is severe NPDR, and 4 is proliferative DR (PDR). Early detection and treatment, particularly in mild to moderate NPDR, may slow DR progression and reduce blindness incidence. Very few papers try to predict the progression of DR using a single CFP \cite{Rom2022,Bora2021}. Despite its difficulty, this task is crucial for better patient follow-up management.

Recently, longitudinal pretext tasks (LPT) have emerged to encode disease progression, such as Longitudinal Self-Supervised Learning (LSSL) introduced by Rivail et al. \cite{Rivail2019} using a Siamese network to predict time lapses between consecutive retinal Optical Coherence Tomography (OCT) scans. Zhao et al. \cite{Zhao2021} proposed a theoretical framework for LSSL using an auto-encoder and an alignment term that forces the topology of the latent space to change in the direction of longitudinal changes. An extension was proposed in \cite{Ouyang} to create a smooth trajectory field and a dynamic graph was computed to connect nearby subjects and enforce maximally aligned progression directions. Authors in \cite{zeghlache} successfully used LSSL in the context of DR to predict the change from grades \{0,1\} to \{2,3,4\}, referred to as Moderate+, between two consecutive CFPs. However, since these approaches are self-supervised, it is unclear whether they learn the disease progression or the longitudinal changes.

Neural Ordinary Differential Equation (NODE) is a new type of neural networks that parameterizes the continuous dynamics of ordinary differential equations. It is ideally suited for solving time-related problems. Time-aware models, including NODEs, have achieved state-of-the-art performance in various tasks related to irregularly-sampled time series data or disease progression \cite{Baytas,Yulia,Gruffaz,zeghlache-prime}. However, NODEs remain challenging to train. Authors in \cite{STEER} proposed a simple yet efficient technique to regularize NODEs by randomly solving ODE for longer time points.

In recent years, mixing augmentation training has been successful in computer vision \cite{mixup,manifold,AutoMix}. Mix-up \cite{mixup} performs on the training set linearly, mixing a random pair of examples and their corresponding labels. Manifold Mix-up \cite{manifold} extends such principle to linear interpolation to the hidden representation. Based on two one-hot labels, the mix-up training generates soft labels that model the relationship between two classes. One-hot labels describe the intra-class relationship, while Mix-up labels describe the inter-class relationship. 

These soft labels modulate the learned decision boundaries, providing the model with more information. In this sense, Mix-up training can be seen as a pretext task. Going further, longitudinal-based Mix-up can be considered as a pretext task that captures the disease's (presumed) linear progression. Motivated by that, we propose a LPT using Manifold Mix-up (MM) \cite{manifold}, which we consider to be more suitable form of Mix-up training for our objective because manifold Mix-up has proven to provide more informative hidden representations. In order to learn feature representations embedded with disease progression that can be reapplied in tackling longitudinal-based problems,  especially for the challenging task of prediction the disease progression based on a single image. We introduce $t_{mix}$ an intermediate time in the course of the disease progression, $t_{mix}$ is the weighted average time between two consecutive examinations. $t_{mix}$ is used to obtain soft labels based on a severity profile. Additionally, it acts as a data augmentation for training time-aware models. To the best of our knowledge, this work is the first to automatically assess DR and forecast its progression using Mix-up training and time-aware models, using only one CFP.

\section{Methods}
\subsection{Preliminary}

\begin{figure}[t]
%
\includegraphics[width=\textwidth]{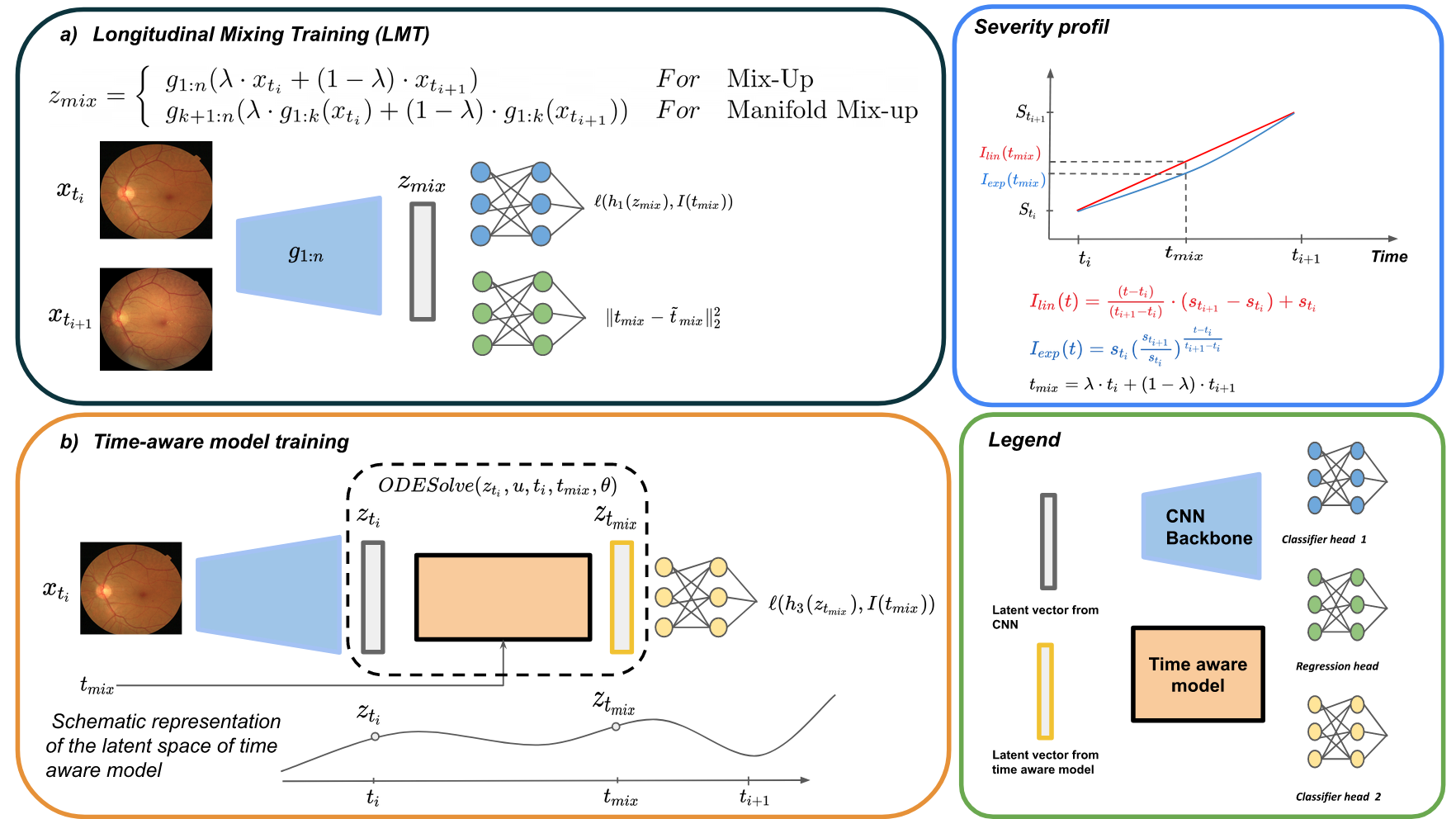}

\caption{Illustration of Longitudinal Mixing Training in a) and time-aware model training using $t_{mix}$ in b).  Fig.\ref{fig:proposed_method}a) and Fig.\ref{fig:proposed_method}.b) can be trained simultaneously or independently.}
\label{fig:proposed_method}
\end{figure}

Let $\mathcal{V}$ be the set of consecutive patient-specific image pairs for the collection of all CFP images. $\mathcal{V}$ contains all $(x_{t_{i}}, x_{t_{i+1}})$ that are from the same patient where $x_{t_{i}}$ is scanned before $x_{t_{i+1}}$ with $i \in [0,m-2]$, $m$ being the number of visits for a given eye. Figure \ref{fig:proposed_method} displays a simplified architecture for clarity. We define backbone $g_{1:n}$ with $n$ layers, where $g_{1:k}$ denotes the part of the neural network mapping the input data to the hidden representation at layer $k$. $h_{l}$ represents a classification or regression head $l$, $\left(y, y'\right)$ one-hot labels, $\text{Beta}\left(\alpha, \alpha\right)$ the Beta distribution and $\ell(.)$ the Binary Cross-Entropy (BCE) loss. We define the mixing operator by $\text{Mix}_\lambda(a, b) = \lambda \cdot a + (1 - \lambda) \cdot b$ with $\lambda \sim \text{Beta}\left(\alpha, \alpha\right)$ where $\lambda \in [0,1]$.

\noindent \textbf{Mix-up} was introduced in \cite{mixup} as a simple regularization method to minimize overfitting in deep neural networks. It linearly interpolates a mini-batch of random examples and their labels to transform the training set. \\
\noindent \textbf{Manifold Mix-up} is an extension of Mix-up to hidden representations \cite{manifold}. During training, a random layer $k$ from a set of eligible layers $S$ in a neural network is selected. It processes two random data mini-batches $\left(x, y\right)$ and $\left(x', y'\right)$, until reaching layer $k$. The Mix-up is then performed on these two intermediate mini-batches $\left(g_{k}\left(x\right), y\right)$ and $\left(g_{k}\left(x'\right), y'\right)$, continuing the forward pass with the mixed representation until the ending layer $n$. These mixed representations are then fed to the classification head $h_{l}$ and projected to the number of classes. 


\noindent \textbf{Neural Ordinary Differential Equations (NODEs)} approximate unknown ordinary differential equations by a neural network \cite{NeuralODE} that parameterizes the continuous dynamics of hidden units $\mathbf{z}\in \mathbb{R}^n$ over time with $\mathbf{t}\in \mathbb{R}$. NODEs are able to model the instantaneous rate of change of $\mathbf{z}$ with respect to $\mathbf{t}$ using a neural network $u$ with parameters $\theta$.

\begin{equation} 
\label{equ:neural_odes}
\lim_{h\rightarrow0}\frac{\mathbf{z}_{t+h}-\mathbf{z}_t}{h}=\frac{d\mathbf{z}}{dh}=u(t,\mathbf{z},\boldsymbol{\theta})
\end{equation} \vspace{-0.1cm} \\

\noindent The analytical solution of Eq.\ref{equ:neural_odes} is given by:
 
\begin{equation} 
\mathbf{z}_{t_{1}} = \mathbf{z}_{t_{0}} + \int_{t_{0}}^{t_{1}}u(t,\mathbf{z},\boldsymbol{\theta})\mathrm{d}t =\textrm{ODESolve}(\mathbf{z}(t_0), u, t_0, t_1, \theta)\label{equ:solveode}
\end{equation}

\noindent where $[t_{0}, t_{1}]$ represents the time horizon for solving the ODE, $u$ being a neural network, and $\theta$ is the trainable parameters of $u$. By using a black-box ODE solver introduced in \cite{NeuralODE}, we can solve the Initial Value Problem (IVP) and calculate the hidden state at any desired time using Eq.\ref{equ:solveode}. We can differentiate the solutions of the ODE solver with respect to the parameters $\theta$, the initial state $\mathbf{z}_{t_0}$ at initial time $t_0$, and the solution at time $t$. This can be achieved by using the adjoint sensitivity method \cite{NeuralODE}. Through the latent representation of a given image, we define an IVP that aims to solve the ODE from $t_{i}$ to a terminal time $t_{i+1}$:
\begin{equation}
    \dot z(t) = u(z(t), t, \theta), \textrm{with the initial value} \:z(t_{i}) = z_{{t_{i}}}
    \label{eq:IVP}
\end{equation}

\subsection{Longitudinal Mixing Training (LMT)}

We denote $s_{t_{i}}$ the severity grade of image $x_{t_{i}}$. $I(t)$ is the severity interpolation function between two consecutive longitudinal pairs. In the linear case, between $x_{t_{i}}$ and $x_{t_{i+1}}$ we have $I_{lin}(t)= \frac{(t-t_{i})}{(t_{i+1}-t_{i})} \cdot (s_{t_{i+1}}-s_{t_{i}}) + s_{t_{i}}$. In conventional Mix-up training, the labels are mixed. Instead, we propose to mix the time between consecutive pairs \textit{$t_{mix} = \text{Mix}_\lambda(t_{i},t_{i+1})$}, then used this $t_{mix}$ to evaluate $I(t)$ and uses this signal as supervision. Motivated by the assumption that the progression of DR is a slow process, we tested another monotonic disease progression profile expressed as $I_{exp}(t) = s_{t_{i}} (\frac{ s_{t_{i+1}}}{s_{t_{i}}})^ \frac{t-t_{i}}{t_{i+1}-t_{i}}$. During training, we one-hot encode the value of interpolation at $t_{mix}$ to get our soft label. Depending on the mix-up method, we have as latent representation of the mixed pair:



\begin{equation}
z_{mix} = \left\{ 
\begin{array}{lll}
g_{1:n}( \text{Mix}_\lambda( x_{t_{i}} ,  x_{t_{i+1}}))  & for&\mbox{Mix-Up}
\\
g_{k+1:n}( \text{Mix}_\lambda( g_{1:k}(x_{t_{i}}),  g_{1:k}(x_{t_{i+1}}))) &for&\mbox{Manifold Mix-up}
\end{array}
\right.
\end{equation} 

During the training of our LMM, we add a \textbf{\textit{time consistency}} loss, as follows:
\begin{equation}
L_{t_{mix}} = \parallel t_{mix} - \tilde{t}_{mix} \parallel_2^2 \label{eq:loss_2}
\end{equation} 

\noindent with $\tilde{t}_{mix} = h_{2}(z_{mix})$, where $h_{2}$ is regression head that predicts the value of the current $ t_{mix}$ for a given pair. This term is inspired by \cite{Rivail2019} and motivated by authors in \cite{mangla2020charting}, who used Manifold Mix-up coupled with SSL loss to enhance the quality of the feature extraction. For the training of the LMM, the total loss is:
\begin{equation}
L= \expectation_{(x_{t_{i}},x_{t_{i+1}}) \sim \mathcal{V}} \expectation_{\lambda \sim \text{Beta}(\alpha, \alpha)}\  \expectation_{k \sim \mathcal{S}}\ell(h_{1}(z_{mix}),I(t_{mix}))+ L_{t_{mix}} \label{eq:Full_equation}
\end{equation}

Concerning the use of $t_{mix}$ for the NODE; instead of solving the ODE from $t_{i}$ to a terminal time $t_{i+1}$ using Eq.\ref{eq:IVP}, we solve to the intermediate time $t_{mix}$ (see Fig.\ref{fig:proposed_method}.b) then use this $t_{mix}$ to evaluate $I(t)$ and take this signal as supervision for training. Note that this approach could be applied to any time-aware model.

\normalsize

\section{Experiments and results}

\noindent \textbf{Dataset.} The proposed models were trained and evaluated on OPHDIAT \cite{ophdiat}, a large CFP database collected from the Ophthalmology Diabetes Telemedicine network consisting of examinations acquired from 101,383 patients between 2004 and 2017. Out of 763,848 interpreted CFP images, 673,017 were assigned a DR severity grade, and the others were non-gradable. Patients range in age from 9 to 91, and image sizes vary from 1440 $\times$ 960 to 3504 $\times$ 2336 pixels. Each examination includes at least two eye images. To limit consecutive pairs without progression, 10412 patients were selected with at least one severity change. Each patient had 2–5 scans, averaging 3.43, spanning an average interval of 4.86 years. This dataset was further divided into training (60$\%$), validation (20$\%$), and test (20$\%$) based on patients. We randomly selected one image per eye for each examination, resulting in 49578 pairs. Our longitudinal downstream task is to predict whether an eye without DR at the initial visit was later graded as having Mild+ DR within two years. 8,111 patients and 13,936 eyes fit this criterion for the training. For the DR assignment, a specific test set consisted of patients assigned the same grade from two ophthalmologists, resulting in 9,734 eyes of 4,996 patients. Except for the registration, we followed the same image processing performed in \cite{zeghlache}. All the timestamp were normalized by 2$\times$365.


\noindent\textbf{Implementation details.} In our basic architecture, we employed a stack of $2$ pre-activated residual blocks (ReLu+BN). In each residual block, the residual feature map was calculated using a series of three 3$\times$3 convolutions, the first of which always halves the number of the feature maps employed at the present scale. Our encoder comprised seven levels; the first six levels are composed of two residual blocks and the latter deals with only one residual block. This provides a final latent representation of size $64\times4\times4$. The last three layers of our backbone were used for the eligible layers S. The different networks were trained for 200 epochs by the AdamW optimizer, OneCycleLR as a scheduler, weight decay of $10^{-4}$, and a batch size of 128, using an NVIDIA A6000 GPU with the PyTorch framework. A grid search was performed for several key hyper-parameters of all Mix-up algorithms and SSL, including $\alpha$ $\in$\{0.2,0.5,1.0,2.0,3.0,5.0,10.0\}, learning rate $\in$ \{$10^{-2}$,$10^{-3}$,$10^{-4}$\} and three different seed values. Concerning the NODE, we used the Pytorch package Torchdiffeq \cite{torchdiffeq}. This library provides ODE solvers, and backpropagation through ODE solutions and a support of the adjoint method for constant memory cost. Our NODE is a combination of dense layers followed by the tanh activation function and the adjoint method with "dopri5" as a solver. The loss in each task was used to monitor the model’s performance on the validation set, the best one was kept. \\ 
\noindent \textbf{Experiments for DR severity assessment using mixing training.} Usually, a new permutation is applied in mixing training at each batch. Since we only use fixed consecutive pairs, the network looks at fewer examples. To perform fair comparisons, we tried multiple permutations that matched the distribution of longitudinal pairs. We report the Quadratic-weighted Kappa for different scenarios of mixing training in Tab.1.\\
\noindent \textbf{Experiments to evaluate the quality of feature extraction.} For the downstream task, we provided the results from both linear evaluation and fine-tuning. The performance was evaluated with the Area Under the receiver operating characteristic Curve (AUC) and reported Tab.2. The linear evaluation was conducted by training a linear layer on top of the pre-trained and frozen encoder and trained using the same set-up that was tried with LMM. For the classical feature extractor, we used AE, VAE \cite{Kingma}, and SimCLR \cite{SIMCLR}. For longitudinal SSL, we used longitudinal Siamese \cite{Rivail2019}, LSSL \cite{Zhao2021}, and LNE \cite{Ouyang}. \\


\hspace{-0.6cm} \begin{minipage}[c]{0.5\textwidth}
\begin{adjustbox}{width=5cm,center}
\label{table:table_comparaison_kappa_mix_vs_long_mix}    
\begin{tabular}{l|c|c|}
      & Kappa & $\alpha$ \tabularnewline
        \hline
        Mix-up $\star$ & 0.7646  & 0.2  \tabularnewline
        Manifold Mix-up $\star$	& 0.7747 & 2.0   \tabularnewline
        \hline
        Mix-up  & 0.7314 	 & 0.2   \tabularnewline
        Manifold Mix-up  & 0.7342 & 2.0  \tabularnewline
        Longitudinal Mix-up (LM) + $I_{lin}(t)$   & 0.7339  & 0.2  \tabularnewline
        Longitudinal Manifold Mix-up (LMM) + $I_{lin}(t)$   & \textbf{0.7511}  &	2.0  \tabularnewline
        Longitudinal Mix-up  (LM) + $I_{exp}(t)$  & 0.5595 &	0.5  \tabularnewline
        Longitudinal Manifold Mix-up (LMM)  + $I_{exp}(t)$  & \textbf{0.7350} 	& 2.0 	 \tabularnewline
    \hline
\end{tabular}
\end{adjustbox}
\captionof{table}{Comparison of the best Kappa for Mix-up training for DR severity assessment.}
\end{minipage}
\begin{minipage}[c]{0.5\textwidth}
\begin{adjustbox}{width=5cm,center}
\centering
\label{table:table_comparaison_AUC_mix_vs_long_mix2}
\begin{tabular}{|l|cc|}
\hline
                  & \multicolumn{2}{l}{AUC ( Mild+ DR within 2 years)} \\ \hline
Weights            & \multicolumn{1}{l|}{Frozen}          & Fine-tuned   \\ \hline
Random            & \multicolumn{1}{l|}{-}               & 0.584        \\ 
MM \cite{manifold} ($\alpha=2.0$)  & \multicolumn{1}{l|}{0.564}           &   0.595      \\ 
AE                & \multicolumn{1}{l|}{0.531}           & 0.569        \\ 
VAE \cite{Kingma} & \multicolumn{1}{l|}{0.510}           & 0.575        \\ 
SimCLR \cite{SIMCLR}  & \multicolumn{1}{l|}{0.544}           & 0.558        \\ 
L-Siamese \cite{Rivail2019} & \multicolumn{1}{l|}{0.562}           & 0.593        \\ 
LSSL \cite{Zhao2021} & \multicolumn{1}{l|}{0.579}           & 0.602        \\ 
LNE \cite{Ouyang}  & \multicolumn{1}{l|}{0.570}           & 0.595   \\

Ours (LMM $\alpha=2.0$) & \multicolumn{1}{l|}{\textbf{0.613}}  & \textbf{0.627} \\ \hline
\end{tabular}
\end{adjustbox}
\label{table:table_comparaison_AUC_mix_vs_long_mix}
\captionof{table}{Results on linear evaluation and fine-tuning of pre-trained model using AUC}
\end{minipage} \\



\noindent \textbf{Experiments using $t_{mix}$ for time-aware models.} Another time-aware model T-LSTM, introduced in \cite{Baytas}, was used for this experiment in order to demonstrate the effectiveness of $t_{mix}$. Both time-aware model take as input the latent representation $z_{t_{i}}$ and the time difference between $x_{t_{i}}$ and $x_{t_{i+1}}$ ($\Delta_{t}$) in order to predict the latent representation of $z_{t_{i+1}}$. For the NODE, it was performed by the mean of IVP defined in Eq.\ref{eq:IVP} while for T-LSTM \cite{Baytas}, the LSTM gates were modulated by $\Delta_{t}$ to produce the future $z_{t_{i+1}}$. We tested three training set-ups:

\begin{enumerate}

    \item Using one image and the time of the next examination, we trained the model presented in Fig.\ref{fig:proposed_method}.b) with the loss $\ell(h_{3}(z_{t_{i+1}},s_{t+1}))$.
    \item  Similarly to (1), we used image $x_{t}$ but instead of giving the time of the next examination, $t_{mix}$ was used and trained with the loss $\ell(h_{3}(z_{t_{mix}},I(t_{mix}))$.
    \item We used (2) with our LMM, i.e., we use (a) and (b) simultaneously (Fig.\ref{fig:proposed_method}). 
\end{enumerate} \vspace{-0.6cm}
\begin{table}[h!]
\begin{adjustbox}{width=6cm,center}

\caption{Comparison of AUCs for the next visit for time-aware model training with and without $t_{mix}$ with linear progression assumption.}
\label{table:table_comparaison_AUC_mix_vs_long_mix2}
\begin{tabular}{l| c| c| c |c }

\cline{1-5} 
& \begin{tabular}[c]{@{}c@{}}AUC\\ Mild+DR\end{tabular} & \begin{tabular}[c]{@{}c@{}}AUC\\ moderate+DR\end{tabular} & \begin{tabular}[c]{@{}c@{}}AUC\\ severe+DR\end{tabular}  & Best $\alpha$ \tabularnewline
\cline{1-5} 

(1) NODE \cite{NeuralODE} & 0.584  &  0.617 & 0.641 & - \tabularnewline
(1) T-LSTM \cite{Baytas} & 0.608 & 0.646 & 0.677 & - \tabularnewline
(2) NODE + $t_{mix}$ (ours) & 0.632  & 0.695 & 0.725 & 2.0 \tabularnewline
(2) T-LSTM + $t_{mix}$  (ours) &  0.610 &   0.661  &  0.725 & 0.5 \tabularnewline
\hline
(3) NODE+LMM  (ours) & \textbf{0.657}  & \textbf{0.721} & \textbf{0.798} & 2.0 \tabularnewline
\hline

\end{tabular}

\end{adjustbox}
\end{table}
The best results for the longitudinal task are obtained with LMM (Tab.2), indicating that it effectively captures disease progression. Moreover, the longitudinal task performed better than classical feature extraction methods, which is aligned with \cite{Emre,Ouyang,Rivail2019,Zhao2021,zeghlache,zeghlache-prime}. Concerning the use of $t_{mix}$, results in Tab.\ref{table:table_comparaison_AUC_mix_vs_long_mix2} indicate that it is beneficial for both time-aware models. We believe $t_{mix}$ plays the role of data augmentation in the context of disease progression for Time-Aware models and is regarded as a method to regularize the training of NODE like \cite{STEER}. The fact that the set-up (3) performs better than other configurations supports the idea that LMM and $t_{mix}$ are beneficial to solve time-related problems.
\begin{figure}[h!]
\RawFloats
\begin{minipage}[c]{0.5\linewidth}
\centering
    \includegraphics[width=0.7\textwidth]{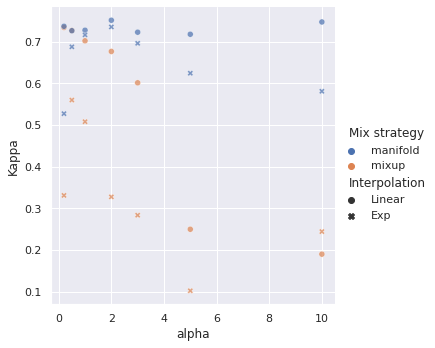}
    \caption{\small{Best value of Kappa when alpha varies for LM and LMM for the two profiles.}}
       \label{fig:sfig1}
\end{minipage}
\begin{minipage}[c]{0.45\linewidth}
    \centering
    \includegraphics[width=0.8\textwidth]{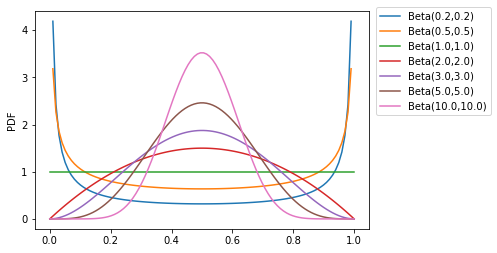}
    \caption{\small{Beta distribution for alpha values used.}}
        \label{fig:sfig2}
\end{minipage}
 \end{figure}

Only LMM adapts $I_{exp}(t)$ (Fig.\ref{fig:sfig1}), indicating that with LMM, other types of severity profiles could be used. We suspect that LM suffers from manifold intrusion \cite{Guo}. Mixing longitudinal pairs create an existing severity grade in the dataset, making training more challenging when using Mix-up. LMM performs better than regular MM, according to Tab.1, when the profile of severity progression was supposed to be linear between two consecutive exams, suggesting that longitudinal pairs are more informative than random pairs with the same label distribution. The more $\alpha$ increases, the more the severity interpolation is taken into consideration because when $\lambda \leq 0.1 \Rightarrow t_{mix} \simeq t_{1}$ and $\lambda = 0.5 \Rightarrow t_{mix} =\frac{t_{0}+t_{1}}{2}$ and finally $ \lambda \geq 0.9 \Rightarrow t_{mix} \simeq t_{0}$, as illustrated by the Beta distribution in Fig.\ref{fig:sfig2}. In Fig.\ref{fig:sfig1} for MML, we found that as alpha increases, Kappa differences between severity profiles increase. This could indicate that $I_{exp}(t)$ is sub-optimal for DR progression. In addition, for $\alpha=10$, the LMM is almost trained with the center of the severity interpolation $I(t)$ as a label and yet, according to Fig.\ref{fig:sfig1}, is able to assign DR with a Kappa of 0.75 (second best value of Kappa in all experiments). This could suggest, like in the original MM \cite{manifold}, that the LMM is able to disentangle factors of variations, such as the one responsible for encoding the disease progression. When alpha is low, there is a higher chance of sampling values of $\lambda$ closer to either 0 or 1 from the Beta distribution. Time-aware models are then practically trained for two tasks: 1- predict the severity grade of the current image ($t_{mix} \simeq t_{i} \Rightarrow I_{lin}(t_{mix}) \simeq s_{t_{i}}$) and 2- predict the next visit grade based on the last exam ($t_{mix} \simeq t_{i+1} \Rightarrow I_{lin}(t_{mix}) \simeq s_{t_{i+1}}$). As a result, the model receives more information, which could explain the increase in performance. However, according to our experiments, T-LSTM does not perform well when $\alpha$ is high. Only the NODE gains from having diverse time point $t_{mix} \in [t_{i},t_{i+1}]$ during training, showing that it can successfully change its hidden dynamic when time varies, in line with the conclusions of \cite{NeuralODE,Yulia}.

\section{Discussion and conclusion}

%

In this paper, we proposed straightforward modifications to Manifold Mix-up training.
This adaptation aims to enhance training of time-aware models for disease progression by introducing $t_{mix}$. The results are encouraging and may help clinicians to choose the best DR screening intervals. Our framework is general and could be easily extended to other Mix-up training \cite{AutoMix,venkataramanan2022alignmixup} and time-aware models \cite{Yulia,debrouwer2019gruodebayes} or to other diseases. However our work has some limitations. We did not register images between consecutive examinations. Image registration is a critical step, as mentioned in \cite{Saha2019}, and could enhances our results. To overcome the lack of grade diversity for a given pair during training, we could train with all potential pairs of a patient in a follow-up, as done in \cite{Emre}. We made a strong assumption on the disease progression by supposing one common severity profile, yet observing good results. Since we can access the full examination, we could use a more accurate interpolation function to better fit the DR progression. We hope this work will benefit the fields of longitudinal analysis and disease progression. \\

\noindent \textbf{Acknowledgments}
The work takes place in the framework of the ANR RHU project Evired. This work benefits from State aid managed by the French National Research Agency under the “Investissement d’Avenir” program bearing the reference ANR-18-RHUS-0008.

%


\bibliographystyle{splncs04}
\bibliography{biblio}
%
%
%

\end{document}